\documentclass[11pt]{article}
\usepackage{ams}
\title{On a two particles system associated to the one spatial
Galilei group }
\author{Joachim Nzotungicimpaye\\Kigali Institute of Education, Department of Mathematics\\ P.O.Box 5039,Kigali-Rwanda\\e-mail kimpaye
@kie.ac.rw~,~nzotungakim@yahoo.fr}
\date{}
\begin{document}
\maketitle
\begin{abstract}
{\it In \cite{perroud} ,M.Daumens and M. Perroud studied a two
Galilean free particles system by realizing the three spatial
dimensional Galilei group on its maximal coadjoint orbit. In this
paper we realize a similar study for the one spatial dimensional
Galilei group . As its maximal coadjoint orbit describes a non free
massive particle \cite{nzo88}, this gives us a two non free galilean
particles system. Use of the barycenter coordinates gives rise to
the notions of a total force and a relative force similar to those
of a total linear momentum and a relative linear momentum
(\cite{perroud}, \cite{goldstein},\cite{davis}). We will distinguish
the case of a non isolated system  from that of an isolated one . We
will the show that the barycenter is accelerated in the first case
while it moves with a constant velocity in the second one.}
\end{abstract}
\section{Introduction}
Let $G$ be a Lie group. As we know \cite{giachetti}, all generic
$G-$hamiltonian spaces are coadjoint orbits of the $G$-action on the
dual of the Lie algebra of the central extension of $G$.\\We also
know \cite{matolocsi} that the projective unitary representations in
quantum mechanics are , in classical mechanics, the symplectic
realizations. In this paper we revisit the one spatial Galilei group
\cite{nzo88}. If $e$ , $p$ and $k$ are respectively the dual of time
translations, space translations and boost,we know
(\cite{nzo88},\cite{nzo94}) that there are three kinds of galilean
systems corresponding each to a coadjoint orbit (a $G-$ homogeneous
symplectic manifold)on which the Galilei group is represented
faithfully. These orbits are:
\begin{itemize}
\item One $O_{(m,f,U)}$ endowed with the symplectic form $\sigma=dp\wedge dq$ and characterized by a mass $m$,
a force $f$ and an internal energy $U=e-\frac{p^2}{2m}+fq$ with
$q=\frac{k}{m}$.
 \item One $O_{(m,U)}$ endowed with the symplectic form $\sigma=dp\wedge dq~~,q=\frac{k}{m}$ and characterized by a mass $m$ and
an internal energy $U=e-\frac{p^2}{2m}$.
\item One $O_{(f,{\it \cal{K}})}$ endowed with the symplectic form
$\sigma=dp\wedge dq~~,q=-\frac{e}{f}$ and characterized by a force
$f$ and an invariant ${\it \cal{K}}=k-\frac{p^2}{2f}$. This is the
result of \cite{nzo88}. In the next section we identify
$O_{(f,\cal{K})}$ with a space-time curved by the force $f$.
\end{itemize}
The second section in this paper recall the galilean momenta on the
three coadjoint orbits. In the third section , similary to the work
\cite{perroud}, we study a two particles systems corresponding to
the orbits $O_{(m_a,f_a,U_a)}~,a=1,2$ (non free massive
particles).The study of a two particles system corresponding to the
orbits $O_{(m_a,U_a)}~,a=1,2$ (free massive particles) does not give
any new result.
\section{Galilean momenta}
Let us recall some useful ingredients on the momentum map and the
apply then them to the Galilean coadjoint orbits . Let $G$ be a Lie
group and let $\cal{G}^*$ be the dual of the Lie algebra $\cal{G}$
of $G$. If $(V,\sigma)$ is a given $G-$symplectic manifold , there
exist a momentum map (\cite{souriau},\cite{smale})$J:V\rightarrow
{\cal{G}}^*$ defined by
\begin{eqnarray}
<J(x),X>=[\lambda(X)](x)
\end{eqnarray}
where $\lambda :{\cal{G}}\rightarrow C^{\infty}(V,\mathbf{R})$ is
the comomentum defined by
\begin{eqnarray}
X\rfloor \sigma=d\lambda(X)
\end{eqnarray}
 The components of the $J(x),~x\in V$ in a given basis of $\cal{G}^*$ are
the classical fundamental observables associated to $G$. If $(X_i)$
is a basis of the Lie algebra $\cal{G}$ whose the dual basis is
$X^{*i}$, then $J(x)=J_i(x)X^{*i} \in \cal{G}^*$ gives rise to
\begin{eqnarray}
J_i(x)=[\lambda(X_i)](x)
\end{eqnarray}
which defines the $i^{th}$ component of the momentum. We will then
denote by $J_X(x)$ the momentum correponding to the generator $X$ of
$\cal{G}$.
\\Let us denote the Galilean generators by $K$ for the boosts, $P$
for the space translations and $E$ for the time translations. From
(\cite{nzo88}, \cite{nzo94}) we verify that
\begin{itemize}
\item the symplectic realization of the Galilei group on
$O_{(m,f,U)}$ is
\begin{eqnarray}\label{cr1}
L_{(x,t,v)}(p,q)=(p-mv+ft,
q+\frac{p}{m}t+\frac{f}{m}\frac{t^2}{2}+x-vt)
\end{eqnarray}
The Galilei Lie algebra is then realized by the hamiltonian vector
fields
\begin{eqnarray}
\rho(K)=m\frac{\partial}{\partial
p}~,~\rho(P)=-\frac{\partial}{\partial
q}~,~\rho(E)=-f\frac{\partial}{\partial
p}-\frac{p}{m}\frac{\partial}{\partial q}
\end{eqnarray}
and the Galilean momentum components are
\begin{eqnarray}\label{momentum1}
J_K(p,q)=mq~,~J_P(p,q)=p~,~J_E(p,q)=\frac{p^2}{2m}-fq
\end{eqnarray}
\item the symplectic realization of the Galilei group on
$O_{(m,U)}$ is
\begin{eqnarray}
L_{(x,t,v)}(p,q)=(p-mv, q+\frac{p}{m}t+x-vt)
\end{eqnarray}
The Galilei Lie algebra is then realized by the hamiltonian vector
fields
\begin{eqnarray}
\rho(K)=m\frac{\partial}{\partial
p}~,~\rho(P)=-\frac{\partial}{\partial
q}~,~\rho(E)=-\frac{p}{m}\frac{\partial}{\partial q}
\end{eqnarray}
and the Galilean momentum components are
\begin{eqnarray}\label{momentum2}
J_K(p,q)=mq~,~J_P(p,q)=p~,~J_E(p,q)=\frac{p^2}{2m}
\end{eqnarray}
\item the symplectic realization of the Galilei group on
$O_{(f,{\it \cal{K}})}$ is
\begin{eqnarray}
L_{(x,t,v)}(\tau,q)=(\tau+t, q+v\tau+x)
\end{eqnarray}
where we have used the non canonical space-time coordinates with
$\tau=\frac{p}{f}$. The Galilei Lie algebra is then realized by the
hamiltonian vector fields
\begin{eqnarray}
\rho(K)=-\tau\frac{\partial}{\partial
q}~,~\rho(P)=-\frac{\partial}{\partial
q}~,~\rho(E)=-\frac{\partial}{\partial \tau}
\end{eqnarray}
and the Galilean momentum components are
\begin{eqnarray}\label{momentum3}
J_K(q,\tau)=f\frac{\tau^2}{2}~,~J_P(q,\tau)=f\tau~,~J_E(p,q)=-fq
\end{eqnarray}
\end{itemize}
Looking to (\ref{momentum1})and(\ref{momentum2})
 permit us to interpret $J_P$ and $J_E$ as
respectively a linear momentum and an energy while the relation
(\ref{momentum3}) tell us that $J_P$ and $J_E$ are respectively an
impulse and a work \cite{goldstein}. Nothing is clear for the
component $J_K$. This is the object of this article. We will do it
by considering the two particles systems .The phase space of this
system will be a cartesian product of two orbits of the same kind.
We will endow it with a symplectic form which is the sum of the
corresponding symplectic forms.
\section{Two galilean particles system}
Following \cite{perroud}, let us consider a symplectic manifold
$(V,\sigma)$ where $V$ is the cartesian product of two orbits
$O_{(m_1,f_1,U_1)}$ and$ O_{(m_2,f_2,U_2)}$  while $\sigma$ is the
sum of the symplectic forms $\sigma_a$ on $O_{(m_a,f_a,U_a)}$ :
\begin{eqnarray}\label{form1}
\sigma=dp_1\wedge dq_1+dp_2\wedge dq_2
\end{eqnarray}
The active action of the Galilei group on $V$ is given by\\
\\
$~~~~~~L_{(x_a,t_a,v_a)}(p_a,q_a)$
\begin{eqnarray}\label{GxG}
=(p_a-m_av_a+f_at_a,
q_a+\frac{p_a}{m_a}t_a+\frac{f_a}{m_a}\frac{t_a^2}{2}+x_a-v_at_a),~a=1,2
\end{eqnarray}
where $v_a$ is the velocity of the particle $a$ with respect an
observer , $(x_a,t_a)$ is the space-time "position" of the particle
$a$ with respect the observer.\\
This means that $V$ describes a two particles system whose the
masses are $m_1$ and $m_2$, while the force acting on
$O_{(m_1,f_1,U_1)}$ is $f_1$, that on $O_{(m_2,f_2,U_2)}$ is $f_2$.
The force $f_1$ ($f_2$) is produced by $O_{(m_2,f_2,U_2)}$
($O_{(m_1,f_1,U_1)}$) for an isolated system or the force $f_1$
($f_2$) is produced by $O_{(m_2,f_2,U_2)}$ ($O_{(m_1,f_1,U_1)}$) and
other objects for an non isolated system.
\subsection{Barycenter decomposition}
Let us introduce the barycenter decomposition $B:V\rightarrow
V_{CM}\times V_{INT}$ where $V_{CM}=\{(p,q)\}$ and
$V_{INT}=\{\pi,\rho \}$ defined by \cite{perroud}
\begin{eqnarray}\label{barycentredecomp}
p=p_1+p_2~,~\pi=\frac{m_2p_1-m_1p_2}{m_1+m_2}~,~q=\frac{m_1q_1+m_2q_2}{m_1+m_2}~,~\rho=q_1-q_2
\end{eqnarray}
where \cite {perroud} $m=m_1+m_2$ is the total mass, $p=p_1+p_2$ is
the total linear momentum, $\pi=\frac{m_2p_1-m_1p_2}{m_1+m_2}$ is
the relative linear momentum while $\rho=q_1-q_2$ is the relative
position. Note that the Galilei group has to transform $(p,q)$ and
$(\pi,\rho)$ according (\ref{cr1}). But we verify that (\ref{GxG})
does not preserve the barycenter decomposition. This will be done by
the subgroup ${\cal{B}}=\{((x_1,t,v_1),(x_2,t,v_2))\subset G
\times G\}$ which act on $(p,q,\pi,\rho)$ according\\
\\
$L_{((x,t,v),(r,t,u))}(p,q,\pi,\rho)$
\begin{eqnarray}\label{barycentreaction}
=(p-mv+ft,q+\frac{p}{m}t+\frac{f}{m}\frac{t^2}{2}+x-vt, \pi-\mu
u+\varphi t,
\rho+\frac{\pi}{\mu}t+\frac{\varphi}{\mu}\frac{t^2}{2}+r-ut)
\end{eqnarray}
where
\begin{eqnarray}
f=f_1+f_2~,~\mu=\frac{m_1m_2}{m_1+m_2}~,~\varphi=\frac{m_2f_1-m_1f_2}{m_1+m_2}
\end{eqnarray}
and
\begin{eqnarray}
r=x_1-x_2~,~x=\frac{m_1x_1+m_2x_2}{m_1+m_2}~,~v=\frac{m_1v_1+m_2v_2}{m_1+m_2}~,~u=v_1-v_2
\end{eqnarray}
We recognize in $\mu$ the reduced mass (\cite{perroud},
\cite{goldstein} ,\cite{davis}). Let us call $f$ and $\varphi$
 the total force and the relative force respectively. The non isolated system is characterized
 by $f\neq 0$ while the isolated one is characterized by $f=0$.\\Note that the
relations defining $\pi$ and $\varphi$ can be rewritten as
 \begin{eqnarray}
 \frac{\pi}{\mu}=\frac{p_1}{m_1}-\frac{p_2}{m_2}~,~\frac{\varphi}{\mu}=\frac{f_1}{m_1}-\frac{f_2}{m_2}
  \end{eqnarray}
  They define the relative velocity $\frac{\pi}{\mu}$ and the relative
  acceleration $\frac{\varphi}{\mu}$.\\
 \subsubsection{Non isolated system}
 It is possible to choose a Galilei frame (the center of
mass frame) in which $p=0, q=0$. The internal Galilei group
$G_{INT}$ for the two-particles system is then the subgroup of
$\cal{B}$ which stabilizes $(p=o,q=0)$ \cite{perroud}. We then
verify that
\begin{eqnarray}
G_{INT}=\{(\frac{f}{m}\frac{t^2}{2}+\frac{m_2}{m}r,t,\frac{f}{m}t+\frac{m_2}{m}u),(\frac{f}{m}\frac{t^2}{2}-\frac{m_1}{m}r,t,\frac{f}{m}t-\frac{m_1}{m}u))\}
\end{eqnarray}
In the barycenter coordinates (\ref{barycentredecomp}) the
symplectic form (\ref{form1}) is
 \begin{eqnarray}
 \sigma=dp\wedge dq+d\pi \wedge d\rho
 \end{eqnarray}
From (\ref{barycentreaction}) we verify that the Lie algebra of
$\cal{B}$ is realized on $V$ by the hamiltonian vector fields
\begin{eqnarray}
P_{CM}=-\frac{\partial}{\partial
q}~,~K_{CM}=m\frac{\partial}{\partial p}
\end{eqnarray}
for the mass center,
\begin{eqnarray}
P_{INT}=-\frac{\partial}{\partial
\rho}~,~K_{INT}=\mu\frac{\partial}{\partial \pi}
\end{eqnarray}
for the fictitious particle and the time translation is realized by
\begin{eqnarray}
E=-\frac{p}{m}\frac{\partial}{\partial q}-f\frac{\partial}{\partial
p}-\frac{\pi}{\mu}\frac{\partial}{\partial
\rho}-\varphi\frac{\partial}{\partial \pi}
\end{eqnarray}
We then verify that the corresponding momenta components are
\begin{eqnarray}\label{mq1}
[J(P_{CM})](p,q;\pi,\rho)=p~,~[J(K_{CM})](p,q;\pi,\rho)=mq
\end{eqnarray}
for the mass center,
\begin{eqnarray}\label{mq1}
[J(P_{INT})](p,q;\pi,\rho)=\pi~,~[J(K_{INT})](p,q;\pi,\rho)=\mu\rho
\end{eqnarray}
for the fictitious particle. Moreover the momentum component
conjugated to time (energy) is
\begin{eqnarray}
 [J(E)](p,q;\pi,\rho)=\frac{p^2}{2m}-fq+\frac{\pi
^2}{2\mu}-\varphi r
\end{eqnarray}
We rewrite it as
\begin{eqnarray}
 J_E(p,q;\pi,r)=T(p,\pi)+V(q,r)
\end{eqnarray}
where $T(p,\pi)=\frac{p^2}{2m}+\frac{\pi ^2}{2\mu}$ is the sum of
the kinetic energies while the second term of the right hand side
,$V(q,r)=-f q -\varphi r$, is the sum of the potential energies.\\We
then see that only the energy component depend on the center of mass
and on the fictitious particle. \\From the fact that (use of
(\ref{barycentreaction}))
\begin{eqnarray}\label{evolution1}
L_{(0,t,0)}(p,q,\pi,\rho)=(p+f
t,q+\frac{p}{m}t+\frac{f}{m}\frac{t^2}{2},\pi+\varphi
t,\rho+\frac{\pi}{\mu}t+\frac{\varphi}{\mu}\frac{t^2}{2})
\end{eqnarray}
 we verify that the motion equations are
\begin{eqnarray}\label{evolutioncm1}
p=m\dot{q}~,~\ddot{q}=\frac{f}{m}
\end{eqnarray}
for the particle in the center of mass and
\begin{eqnarray}\label{evolutionfictious}
\pi=\mu\dot{\rho}~,~\ddot{\rho}=\frac{\varphi}{\mu}
\end{eqnarray}
for the fictitious particle.
   Note also that from
(\ref{barycentredecomp}) and (\ref{mq1}), we write
\begin{eqnarray}
\frac{[J(K_{CM})](p,q,\pi,\rho)}{m}=\frac{m_1q_1+m_2q_2}{m_1+m_2}
\end{eqnarray}
which relates the momentum component $[J(K_{CM})](p,q,\pi,\rho)$ to
the center of mass position $q$. We then see that the center of mass
is accelerated by the total force $f$, the acceleration being
$a=\frac{f}{m}$ while the fictitious particle is also accelerated by
the relative force $\varphi$ , the acceleration being
$\gamma=\frac{\varphi}{\mu}$.
\subsubsection{Isolated system}
In this case, $f_1=-f_2=\varphi$. The internal Galilei group is
\begin{eqnarray}
G_{INT}=\{(\frac{m_2}{m}r,t,\frac{m_2}{m}u),(-\frac{m_1}{m}r,t,-\frac{m_1}{m}u))\}
\end{eqnarray}
which is similar to that of \cite{perroud}. Moreover the the action
of the group $\cal{B}$ is\\
\\
$L_{(x,t,v),(r,t,u)}(p,q,\pi,\rho)$
\begin{eqnarray}\label{barycentreaction1}
=(p-mv,q+\frac{p}{m}t+x-vt, \pi-\mu u+\varphi t,
\rho+\frac{\pi}{\mu}t+\frac{\varphi}{\mu}\frac{t^2}{2}+r-ut)
\end{eqnarray}
The only changed momentum component is the energy which becomes
\begin{eqnarray}\label{energy}
 [J(E)](p,q;\pi,\rho)=\frac{p^2}{2m}+\frac{\pi
^2}{2\mu}-\varphi \rho
\end{eqnarray}
Moreover the motion equations are now
\begin{eqnarray}\label{evolutioncm1}
p=m\dot{q}~,~\ddot{q}=0
\end{eqnarray}
for the particle in the center of mass,
\begin{eqnarray}\label{evolutionfictious}
\pi=\mu\dot{r}~,~\ddot{\rho}=\frac{\varphi}{\mu}
\end{eqnarray}
for the fictitious particle. They tell us that the barycenter moves
on a straight line with a constant velocity $\frac{p}{m}$ as it is
usual (\cite{goldstein},\cite{davis}) while the fictitious particle
is still accelerated by the relative force $\varphi=f_1=-f_2$ like
in the non isolated case.


\begin{thebibliography}{99}
\bibitem{giachetti}Giachetti, R. {\it Rev. Nuovo Cimento}
$\textbf{4}(12),1-63, 1981$
\bibitem{matolocsi}Matolocsi, T., {\it Ann.Univ.Sci.Bud.sect.Math},$\textbf{20},71-85,1977$
\bibitem{souriau}J.M.Souriau,{\it Structure des syst\`{e}mes
Dynamiques},Dunod,$1970$
\bibitem{smale}S.Smale, Inv.Math,$10$,,$305-331$,$1970$
\bibitem{lev1}J.M.Levy-Leblond, {\it Journal of Mathematical
Physics}, {\bf $9$}, $1605-, 1968$
\bibitem{nzo88}J.Nzotungicimpaye, {\it Galilei-Newton law by symplectic realizations}, Lett. Math.Phys. $15, 101-110,1988$
\bibitem{nzo94}J.Nzotungicimpaye, {\it Jertk by group theoretical
methods}, J.Phys. A:Math.Gen. $27, 4519-4526 , 1994$
\bibitem{perroud}M.Daumens and M.Perroud, {\it Internal Galilei group for a two-particle system},J.Math.Phys. ,$20, 1077-,1979$
\bibitem{goldstein}H.Goldstein, {\it Classical Mechanics},Addison-Wesley Publishing Company, $1980$
\bibitem{davis}A.A.Davis,{\it Classical Mechanics},Harcourt Brace Jovanovich Publishers ,$1986$
\end{thebibliography}
\end{document}